\documentclass[twoside,showpacs,twocolumn]{revtex4}%
\usepackage{amsmath}
\usepackage{epsfig}
\usepackage{amsfonts}
\usepackage{amssymb}
\usepackage{graphicx}%
\newcommand{\be}{\begin{equation}}
\newcommand{\ee}{\end{equation}}
\newcommand{\bea}{\begin{eqnarray}}
\newcommand{\eea}{\end{eqnarray}}

\begin{document}
\title{Pulsed and continuous measurements of exponentially decaying systems}
\author{Francesco Giacosa$^{\text{(a)}}$ and Giuseppe Pagliara$^{\text{(b)}}$}
\affiliation{$^{\text{(a)}}$Institut f\"{u}r Theoretische Physik, J. W. Goethe
Universit\"{a}t, Max-von-Laue-Str. 1, 60438 Frankfurt, Germany, $^{\text{(b)}%
}$Dip.~di Fisica e Scienze della Terra dell'Universit\`{a} di Ferrara \\
and INFN Sez.~di Ferrara, Via Saragat 1, I-44122 Ferrara, Italy }

\begin{abstract}
We study the influence of a detector on the decay law of a quantum state whose
\textquotedblleft undisturbed\textquotedblright\ survival probability is
purely exponential. In particular, we consider a detector with a finite energy
band of detection, i.e. it interacts only with decay products having an
energy within a certain range of values. In one case, we assume that the
detector performs many repeated measurements at short time intervals in all of
which a collapse of the wave function occurs (bang-bang or pulsed-type
measurements). In the second case, we assume a continuous measurement which
preserves unitarity. We confirm the slowing down of the decay in presence of a
measuring apparatus, the Quantum Zeno effect, but the outcomes of the detector
are in general qualitatively and quantitatively different in the two cases. In
turn, this implies that the so-called Schulman relation (the equivalence of
pulsed and continuous measurements) does not hold in general and that it is in
principle possible to experimentally access how a certain detector performs a
measurement.

\end{abstract}

\maketitle

\emph{Introduction: }The quantum Zeno effect is a phenomenon which consists in
a slowing down of the decay law due to the influence of the measurements
\cite{misra,dega,Koshino:2004rw}. It was first observed as an inhibition of a
quantum transition between two energy levels in Ref. \cite{itano} and later on
confirmed in Ref. \cite{balzer}. These experiments can be interpreted as the
effect of a pulsed series of ideal measurements at time intervals $\tau$,
where each measurement causes the collapse of the wave function.

When the quantum system is unstable (i.e. it couples to a continuum), the
occurrence of the quantum Zeno and anti-Zeno effects are related to the
non-exponential behavior of the survival probability at short times after the
preparation of the initial unstable state, see e.g. Refs.
\cite{ghirardi,facchiprl,kurizki}. In turn, these deviations are regulated by
the existence of a high-energy scale (alias a `cutoff') $\Lambda$ in the
spectral function of the unstable system which depends on its interaction with
the decay products. If the time interval $\tau$ between two measurements is of
the order of (or smaller than) $1/\Lambda$ the decay probability is
effectively modified \cite{gp}. Such effects could be seen experimentally by
using tailored unstable systems in the pioneering works of Refs.
\cite{raizen1,raizen2} (also in this case, the collapse postulate can be used
to understand the results). However, in natural unstable nonrelativistic
quantum systems, such as an excited atomic state, $\Lambda$ is usually very
large (much larger than the decay width) and not experimentally accessible
\cite{pascaatomico}. Conversely, in the high-energy realm of strong decaying
particles such a cutoff is comparable with the decay width but the
corresponding lifetime is too short to allow a measurement of the temporal
evolution of the system \cite{zenoqft,duecan}.

The wave function's collapse postulate is \emph{not} the only way to describe
the effect of a measurement: a different description makes use of a unitary
evolution arising from the interaction of the quantum state with the detector
\cite{Koshino:2004rw,schulman,pascapulsed}. In the theoretical work of
Schulman \cite{schulman}, it was shown how a pulsed series of measurements at
time intervals $\tau$, in all of which the collapse of the wave function
occurs, is analogous to a continuous measurement with detection's efficiency
$\sigma$, in which a unitary evolution is implemented and no collapse of the
wave function takes place. The equivalence of the two types of measurements is
expressed in the so-called Schulman relation $\tau=4/\sigma$ which holds in
the limit $\tau\rightarrow0$. Indeed, a continuous measurement of a quantum
transition and the validity of the Schulman relation were experimentally
confirmed for a two-level system in Ref. \cite{ketterle}.

In Ref. \cite{Koshino:prl} it was realized that the quantum Zeno effect can
take place also in exponentially decaying systems ($\Lambda\rightarrow\infty
$), provided that the experimental apparatus has a finite energy band of
detection. In turn, this opens the exciting possibility of modifying the
effective decay rates even in the case of natural unstable systems (atoms,
nuclei, etc.). Here we reconsider this possibility by extending the results of
Ref. \cite{Koshino:prl}: we will show in particular that in the case of a experimental apparatus
with a finite bandwidth also the pulsed measurement leads to the Quantum Zeno effect 
and that the Schulman relation does \emph{not}
hold. The two types of measurement provide actually different
`no-click' probabilities (this is the probability that the detector does not
make click up to the instant $t$). This result implies that it is in principle
possible to distinguish between the two types of measuring procedures and thus
to also test the assumption of the collapse of the wave function as a ``real''
physical phenomenon \cite{bassi,bassighirardi,grw,penrose,diosi}.

\emph{Pulsed (bang-bang) measurement: }We consider an unstable state
$\left\vert S\right\rangle $ which couples to a continuum of states
$\left\vert k\right\rangle $ through the following Lee Hamiltonian
$H=H_{0}+H_{1}$ \cite{lee,giacosapra}:%
\begin{equation}
H_{0}=M_{0}\left\vert S\right\rangle \left\langle S\right\vert +\int_{-\infty
}^{+\infty}dk\omega(k)\left\vert k\right\rangle \left\langle k\right\vert
\text{ , } \label{h0}%
\end{equation}%
\begin{equation}
H_{1}=\int_{-\infty}^{+\infty}dk\frac{gf(k)}{\sqrt{2\pi}}\left(  \left\vert
S\right\rangle \left\langle k\right\vert +\text{h.c.}\right)  \text{ ,}
\label{hint}%
\end{equation}
where $M_{0}$ and $\omega(k)$ are the energy eigenvalues of the (unperturbed)
Hamiltonian $H_{0}$. With no loss of generality we set $M_{0}=0.$ Moreover, we
assume that $\omega(k)=k$: the energy is not bounded from below. Although
unphysical, this is usually a very good approximation because the low energy
threshold is typically very far away from $M_{0}$. The parameter $g$ is the
coupling strength and $f(k)$ is a dimensionless form factor which regulates
the coupling of the unstable state with each state of the continuum. In this
work we make the very simple choice $f(k)=1$ (the unstable state couples with
all the decay products with the same strength, i.e. $\Lambda
\rightarrow\infty$).

With these choices, the time evolution reads (see Refs.
\cite{giacosapra,duecan} for details)
\begin{equation}
e^{-iHt}\left\vert S\right\rangle =a(t)\left\vert S\right\rangle
+\int_{-\infty}^{+\infty}dkb(k,t)\left\vert k\right\rangle
\end{equation}
where
\[
a(t)=e^{-\Gamma t/2}\text{ , }\Gamma=g^{2}\text{ , }b(k,t)=\sqrt{\frac{\Gamma
}{2\pi}}\frac{e^{-ikt}-e^{-\Gamma t/2}}{k+i\Gamma/2}\text{ .}%
\]
The survival probability, i.e. the probability that the state $\left\vert
S\right\rangle $ prepared at $t=0$ is still in its initial state at the
instant $t$, is given by $p(t)=\left\vert a(t)\right\vert ^{2}=e^{-\Gamma t}$,
the usual exponential decay law \cite{ww,scully,ford}. Note, if we make a
series of measurements of the unstable state at time intervals $\tau$, after
$n$ steps (alias, at the time $t=n\tau$) the no-click probability is given by
$p(\tau)^{n}=p(t).$ As expected, in the exponential limit one does not have
the Zeno effect: one single measurement at the time $t$ gives the same outcome
of a series of measurements. However, this is true only when a measurement of
the initial state $\left\vert S\right\rangle $ can be exactly performed.

In the following, we assume to make a different type of measurement: our
hypothetical detector detects the final state $\left\vert k\right\rangle $
(such as the photon in an atomic spontaneous emission) \emph{only} in the
energy range $(-\lambda,\lambda)$. The function
\begin{equation}
w_{\lambda}(t)=\int_{-\lambda}^{+\lambda}\mathrm{dk}\left\vert
b(k,t)\right\vert ^{2}%
\end{equation}
\ plays an important role and has a clear interpretation: if we perform a
single, ideal measurement at the instant $\tau$, the probability that the
detector makes `click' (meaning that the photon is detected) is given by
$w_{\lambda}(\tau)$. Now, let us make a series of $n$ measurements at times
$\tau,$ $2\tau,$ ..., $t=n\tau$: the probability that the detector has
\emph{not} detected the decay product(s) at the instant $t$ is given by:
\begin{equation}
p_{\text{no-click}}^{BB}(t)=1-w_{\lambda}(\tau)\frac{1-e^{-\Gamma t}%
}{1-e^{-\Gamma\tau}}\text{ }\label{pncbb1}%
\end{equation}
where the suffix $BB$ stands for bang-bang. The proof of Eq. (\ref{pncbb1}) is
obtained by taking into account the collapse at each ideal measurement, which
leads to $p_{\text{click}}^{BB}(n\tau)=w_{\lambda}(\tau)p((n-1)\tau)$ where
$p_{\text{click}}^{BB}(n\tau)$ is the probability that the detector makes
click at the $n$-th measurement. The quantity $p_{\text{no-click}}^{BB}(t)$ is
always an exponential function for any value of $\lambda$ and $\tau.$ In the
limit $t\rightarrow\infty$,
\begin{equation}
p_{\text{no-click}}^{BB}(\infty)=1-w_{\lambda}(\tau)/(1-e^{-\Gamma\tau})\text{
.}\label{pncbbinf}%
\end{equation}
There is a nonzero probability that the detector will never make click, which
depends both on the range $\lambda$ and the time interval $\tau$. One can
regard this saturation phenomenon of the survival probability as a
manifestation of the Quantum Zeno dynamics \cite{qzd} which is related to our
simplifying assumption of zero efficiency of the detector for values of energy
outside the bandwidth.

\begin{figure}[ptb]
\vskip0.5cm \begin{centering}
\epsfig{file=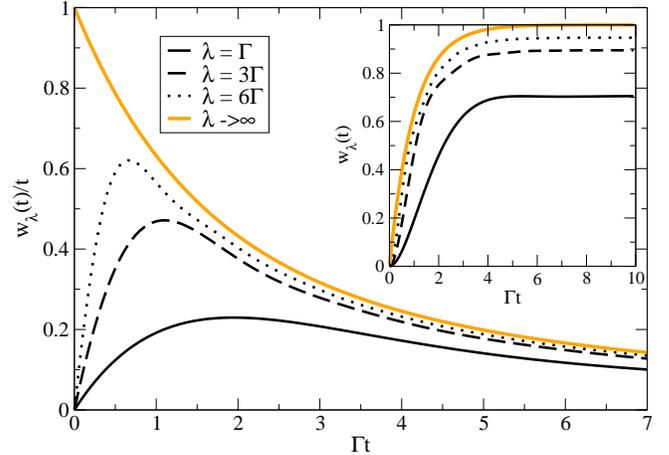,height=8.5cm,width=6cm,angle=-90}
\caption{(Color on-line) The ratio between the decay probability $w_{\lambda}(t)$ and $t$
is displayed as a function of time for different values of $\lambda$. Only in the limit
$\lambda \rightarrow \infty$ the decay probability $w_{\lambda}(t)$ is linear in time while for finite values of $\lambda$ it shows
a quadratic behaviour at short times. Note, the limits $\lambda\rightarrow\infty$
and $t\rightarrow 0$ do not commute. In the insert we show $w_{\lambda}(t)$, notice that for $t\rightarrow\infty$ it saturates to a value smaller than one if $\lambda$ is finite.}
\end{centering}\end{figure}The function $w_{\lambda}(t)$ is plotted in Fig. 1.
At small times, $t\lesssim\lambda^{-1}$, it grows \textit{quadratically with
$t$}. Thus, when a series of measurement is performed with $\tau\rightarrow0$
the quantity $w_{\lambda}(\tau)/(1-e^{-\Gamma\tau})\rightarrow0$ , and
$p_{\text{no-click}}^{\text{BB}}(t)\rightarrow1$: quite remarkably, we obtain
a (full) Zeno effect induced by pulsed measurements in a system whose
unperturbed decay law is exponential. For $\lambda\rightarrow\infty$ one has
$w_{\lambda}(t)=1-e^{-\Gamma t}$ (it grows now \textit{linearly with $t$} at
small times), implying that $p_{\text{no-click}}^{BB}(t)=p(t)=e^{-\Gamma t}.$
No Zeno-effect induced by the detector takes place in this limit and the usual
exponential survival probability is measured, as expected.

\emph{Continuous measurement and comparison: }A simple and effective way to
model the continuous measurement is to add to the Lee Hamiltonian of Eq.
(\ref{h0}) a term which couples the final states $\left\vert k\right\rangle $
to corresponding states of the measurement apparatus. In turn, this coupling
makes the decay products also unstable by effectively modifying their energy
eigenvalues as
\begin{equation}
\omega(k)=k\rightarrow k-i\sigma(k)/2\text{ ,} \label{substitution}%
\end{equation}
where $\sigma(k)$ parametrizes the efficiency of the measuring apparatus
\cite{pascapulsed} and expresses that fact that a real experimental apparatus
is more efficient in detecting certain values of the energy of the decay
products [a neutrino detector for instance is usually based on inelastic weak
interactions of neutrinos with the material of the detector: low energy
neutrinos (much below $1$ MeV) basically cannot be detected]. In order to make
contact with the bang-bang case, we assume that (again setting $M_{0}=0$):
\begin{equation}
\sigma(k)=\sigma\text{ for }k\subset(-\lambda,\lambda)\text{, }0\text{
otherwise.} \label{sigmaofk}%
\end{equation}
Thus, the detector can make click only if the energy $k$ of the outgoing
particle lies within the range $(-\lambda,\lambda)$, while for all the other
values there is no coupling with the apparatus. Intuitively, $\sigma^{-1}$ is
analogous to the time interval $\tau$ of the pulsed measurements. Indeed, in
the case of intrinsic deviations from the exponential law (such as the
existence of a finite cutoff $\Lambda$ encoded in a non constant form factor
$f(k)$), the Schulman relation $\tau=4/\sigma$ holds \cite{schulman}. As we
shall see, in the finite-bandwidth detectors analyzed in this paper this
relation is \emph{not} valid.

\begin{figure}[ptb]
\vskip0.5cm \begin{centering}
\epsfig{file=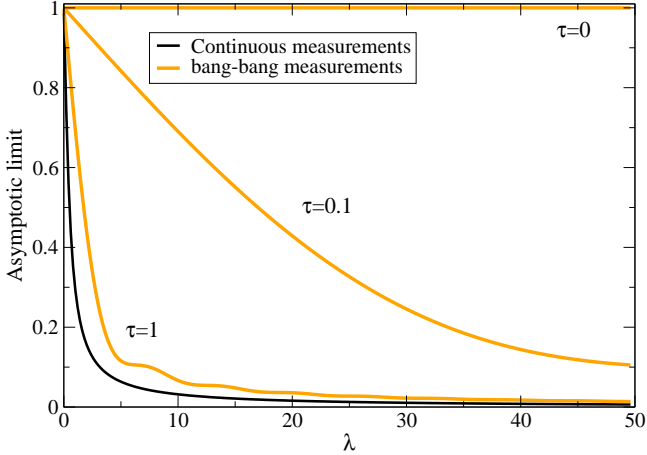,height=8.5cm,width=6cm,angle=-90}
\caption{(Color on-line) We display the saturation value of the no-click
probability in the continuous case (r.h.s. of Eq. (14), valid in the limit of large $\sigma$) as a function of $\lambda$. We also show the 
bang-bang case (see Eq. (6)) 
as a function of $\lambda$ and for different values of $\tau$.}
\end{centering}\end{figure}

The Hamiltonian $H^{C}$ obtained via Eqs. (\ref{h0}) and (\ref{sigmaofk}) upon
the replacement (\ref{substitution}) is not Hermitian (it has complex
eigenvalues) but this does not represent a problem. In principle, the full
evolution makes use of the full Hermitian Hamiltonian $H_{full}$ that involves
the state $\left\vert D_{0}\right\rangle ,$ which describes the detector which
has \emph{not} measured the decayed particle, and $\left\vert D_{k}^{\alpha
}\right\rangle $, which describes one of the (many possible, $\alpha$
enumerates them) states corresponding to the detector which has measured the
decay product with energy $k$:%

\begin{align}
e^{-iH_{full}t}\left\vert S\right\rangle \left\vert D_{0}\right\rangle  &
=\left(  e^{-iH^{C}t}\left\vert S\right\rangle \right)  \left\vert
D_{0}\right\rangle +\nonumber\\
&  \int_{-\infty}^{+\infty}\mathrm{dk}\sum_{\alpha}q(k,t)\left\vert
k\right\rangle \left\vert D_{k}^{\alpha}\right\rangle \text{ .}
\label{apparatus}%
\end{align}
Thus, for the issues we want to discuss here, it is enough to concentrate on
the subspace of the detector $\left\vert D_{0}\right\rangle $. The temporal
evolution reads $e^{-iH^{C}t}\left\vert S\right\rangle =a^{C}(t)\left\vert
S\right\rangle +\int_{-\infty}^{+\infty}dkb^{C}(k,t)\left\vert k\right\rangle
,$ where the suffix $C$ stands for continuous measurement. The survival
probability is
\begin{equation}
p^{C}(t)=\left\vert a^{C}(t)\right\vert ^{2}=\int_{-\infty}^{+\infty
}\mathrm{dE}d_{S}(E)e^{-iEt}\text{ ,}%
\end{equation}
where $d_{S}(E)=$ $\frac{-1}{\pi}\operatorname{Im}G_{S}(E)$ is the energy
distribution and
\[
G_{S}(E)=\left\langle S\right\vert \left[  E-H^{C}+i\varepsilon\right]
^{-1}\left\vert S\right\rangle =\left[  E+\Sigma(E)+i\varepsilon\right]  ^{-1}%
\]
is the propagator of the unstable state $\left\vert S\right\rangle .$ The
self-energy $\Sigma(E)$ reads
\begin{equation}
\Sigma(E)=\frac{g^{2}}{2\pi}\log\left(  \frac{\lambda+E}{\lambda-E}\right)
+\frac{g^{2}}{2\pi}\log\left(  \frac{E-\lambda+i\sigma/2}{E+\lambda+i\sigma
/2}\right)  \text{ } \label{self}%
\end{equation}
and depends on the parameters $\lambda$ and $\sigma$ which specify the
detector intrinsic bandwidth and efficiency. Finally:%

\begin{equation}
b^{C}(k,t)=\sqrt{\frac{\Gamma}{2\pi}}\int_{-\infty}^{+\infty}\mathrm{dE}%
d_{S}(E)\left(  \frac{e^{-i\omega(k)t}-e^{-iEt}}{\omega(k)-E}\right)  \text{
.}\label{b}%
\end{equation}
Just as in the bang-bang case, we ask which is the probability that the
detector does \emph{not} fire up at the instant $t$. The resulting no-click
probability at the time $t$ is given by
\begin{equation}
p_{\text{no-click}}^{C}(t)=p^{C}(t)+w^{C}(t)\text{ ,}%
\end{equation}
where $w^{C}(t)=\int_{-\infty}^{+\infty}\mathrm{dk}\left\vert b^{C}%
(k,t)\right\vert ^{2}$is the probability that the unstable state has decayed
but the detector did not register it. An important quantity is
$p_{\text{no-click}}^{C}(\infty),$ which describes the probability that the
detector does not make click even if one waits a very long time. The analytic
expression of this quantity reads:%
\begin{equation}
p_{\text{no-click}}^{C}(\infty)=\frac{g^{2}}{\pi}\int_{\lambda}^{\infty
}dk\left\vert k+\Sigma(k)\right\vert ^{-2}\text{ }\simeq1-w_{\lambda}%
(\infty)\text{ ,}\label{pnccinf}%
\end{equation}
where the r.h.s. of the equation becomes a very good approximation for
increasing $\sigma.$ $p_{\text{no-click}}^{C}(\infty)$ is only logarithmically
dependent on $\sigma$ and for large but finite values of $\sigma$ approaches 0
due to the fact that $w_{\lambda}(\infty)\sim1$ (see insert in Fig. 1). These
properties are utterly different from the asymptotic behavior
$p_{\text{no-click}}^{BB}(\infty)$ (Eq. (\ref{pncbbinf})) which strongly
depends on $\tau$ for small $\tau$ as shown in Fig.2 (orange lines).

In the following we discuss some relevant cases which we also compare to the
bang-bang results:

(a) A trivial limit holds for $\sigma=0$, for which the time reaction of the
detector is infinite (no coupling with the detector). Being in this case the
Hamiltonian $H^{C}$ Hermitian, we obtain $p_{\text{no-click}}^{C}%
(t)=p^{C}(t)+w^{C}(t)=1$. There is obviously never a click. (This scenario
corresponds to the limit $\tau\rightarrow\infty$ in the bang-bang case:
clearly, there is no click because no measurement takes place.)

(b) For $\lambda\gg\Gamma$ and $\lambda\gg\sigma$ the self-energy reads
$\Sigma(E)=i\Gamma/2,$ which implies $p^{C}(t)=e^{-\Gamma t}$. The no-click
probability reads:%
\begin{equation}
p_{\text{no-click}}^{C}(t)=e^{-\Gamma t}+\frac{\Gamma}{\Gamma-\sigma}\left(
e^{-\sigma t}-e^{-\Gamma t}\right)  \text{ ,} \label{largelambda}%
\end{equation}
which is manifestly different from the projective measurements, where for
large $\lambda$ one has $p_{\text{no-click}}^{BB}(t)=e^{-\Gamma t}$
(independently on the value of $\tau$). Indeed, the present case applies to
the parameter choices studied numerically in\ Ref. \cite{Koshino:prl}. Only if
we \emph{additionally} require that $\sigma\gg\Gamma$, Eq. (\ref{largelambda})
reduces to the usual exponential law $p_{\text{no-click}}^{C}(t)=e^{-\Gamma
t}$.

\begin{figure}[ptb]
\vskip 0.5cm \begin{centering}
\epsfig{file=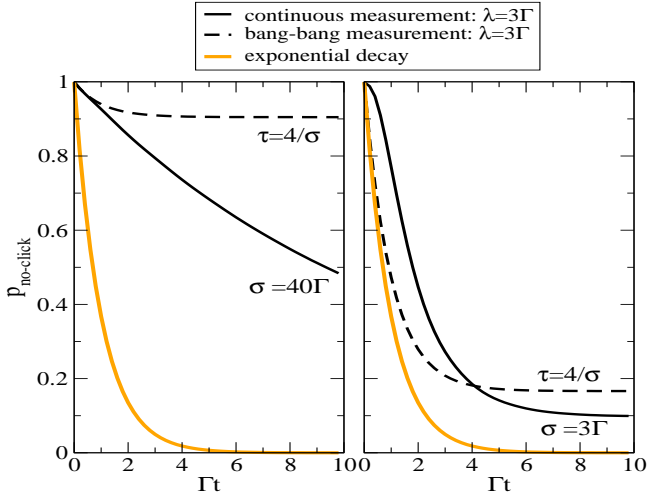,height=8.5cm,width=6.5cm,angle=-90}
\caption{(Color on-line) $p_{\text{no-click}}$ as a function of time for continuous and bang-bang measurements for two values of the detection efficiency $\sigma$ and their relative values of $\tau$ as prescribed by the Schulman rule; also the exponential decay is shown for comparison. The results of the two types of measurements are qualitatively different, the difference becoming larger
for larger values of $\sigma$. }
\end{centering}
\end{figure}

(c) We study numerically the case of finite time response (finite $\sigma$).
We set again $\lambda=3\Gamma$ and display $p_{\text{no-click}}^{C}(t)$ for
$\sigma=40\Gamma$ and $\sigma=\lambda=3\Gamma$ (left and right panels of Fig.
3 respectively). We also show for comparison the bang-bang function
$p_{\text{no-click}}^{BB}(t)$ (Eq. (\ref{pncbb1})) for the Schulman's value
$\tau=4/\sigma$. In the left panel the value of $\sigma$ is large: for short
values of $t$ the two curves lie on top of each other, as the Schulman
relation prescribes. However, they soon depart from each other and, as
explained above, they have utterly different asymptotic values:
$p_{\text{no-click}}^{C}(\infty)<p_{\text{no-click}}^{BB}(\infty)$. If small
values of $\sigma$ are considered ($\sigma=3\Gamma$ see right panel), the
Schulman prescription fails both at short and at long times. Notice that, by
varying the choice of $\tau$ does not improve the correspondence, because it
either spoils the short-time or the long-time agreement. In conclusion, in the
case of finite bandwidth detectors, the Schulman relation (i.e. the
equivalence between bang-bang and continuous measurements) is \emph{not} valid.

Quite interestingly, for large $\sigma$ the function $p_{\text{no-click}}%
^{C}(t)$ can be well approximated by%
\begin{equation}
p_{\text{no-click}}^{C}(t)\simeq\left(  1-p_{\text{no-click}}^{C}%
(\infty)\right)  e^{-\tilde{\Gamma}t}+p_{\text{no-click}}^{C}(\infty
)\label{pnccappr}%
\end{equation}
where the effective decay width reads $\tilde{\Gamma}=-i\frac{\Gamma}{\pi}%
\ln\left(  \frac{-\lambda+i\sigma/2}{\lambda+i\sigma/2}\right)  $. The
quantity $\tilde{\Gamma}$ as function of $\sigma$ equals $\Gamma$ for
$\sigma=0,$ then it decreases monotonically and and approaches zero for
$\sigma\rightarrow\infty$ (Zeno effect). The expression (\ref{pnccappr}) shows
that for large $\sigma$ the asymptotic value is approached very slowly.

\emph{Discussions and conclusions: }We have calculated the probability that a
detector measuring the decay product(s) in a certain finite bandwidth does
\emph{not} make click. We find that there is clear and, in principle
measurable, difference between the pulsed (non-unitary) and continuous
(unitary) time evolution of the system.

A crucial ingredient of the continuous measurement is the function
$\sigma(k),$ where $1/\sigma(k)$ is the time response of the detector when $k$
is the energy of the measured decay product. Here we have used the very simple
expression of Eq. (\ref{sigmaofk}). An interesting outlook is to study the
continuous measurement for more complicated form of $\sigma(k),$ see also
Refs. \cite{koshinorelation,koshinofalse,goto}. In particular, as shown in
Ref. \cite{koshinofalse}, not only the Zeno effect but also the anti-Zeno
effect can be induced by the measurement. Moreover, our discussion is not
restricted to quantum mechanics only, but is valid, with minor changes, in a
relativist quantum field theoretical context \cite{duecan,zenoqft} and it is
thus potentially applicable to decays of fundamental particles as well.

Finally, an interesting implication of our results is the possibility to
distinguish whether the experimental apparatus performs a continuous
measurement or not by tuning its detection efficiency. The most practical
choice is to measure the same unstable system with different apparata which
will have in general quite different efficiencies (which would affect the
parameters $\tau$ and $\sigma$ ). If the asymptotic value of the no-click
probability depends strongly on the choice of the experimental apparatus, one
could deduce that the bang-bang model is correct. On the other hand, in the
case of a continuous measurement case the asymptotic value of the no-click
probability has a very mild (logarithmic) dependence on the detection
efficiency. We regard this argument as a \textquotedblleft proof of
principle\textquotedblright\ of the possibility to test alternatives of QM in
which the collapse of the wave function is a real physical phenomenon which
takes place when the involved number of particles is large enough, as it is
surely the case for a macroscopic detector
\cite{bassi,bassighirardi,grw,penrose,diosi}. Conversely, in the case of
continuous measurements, unitary evolution always holds and that the wave
function never collapses \cite{pascacollapse}. A possibility is to interpret
this unitary evolution as a split in many branches (the so-called Everett
interpretation \cite{everett}), which do not interact with each other because
of decoherence \cite{schlosshauer}. A vivid ongoing discussion about
fundaments of QM concerns the emergence of probabilities in such a context: at
present it is not clear if one is able to derive the Born rule \cite{dsw} or
not \cite{rae}. In conclusion, our analysis shows that theoretical studies and
dedicated experiments on unstable systems, besides the possible technological
applications, might also help to address fundamental issues in quantum mechanics.

\bigskip

\textbf{Acknowledgments: }the authors thank F. Sauli and T. Wolkanowski-Gans
for useful discussions. G.P. acknowledges financial support from the Italian
Ministry of Research through the program \textquotedblleft Rita Levi
Montalcini\textquotedblright.

\end{document}